\documentstyle[preprint,aps]{revtex}

 \draft

\begin{document}

 \title{ A SUPERFLUID GYROSCOPE WITH COLD ATOMIC GASES}
 
 \author{S. Stringari}
\address{Dipartimento di Fisica, Universit\`{a} di Trento,}
\address{and Istituto Nazionale per la Fisica della Materia,}
\address{I-38050 Povo, Italy}

  \date{\today}

  \maketitle
\begin{abstract}
\noindent
 A trapped  Bose-Einstein condensed atomic gas containing a quantized 
 vortex 
 is predicted to exhibit precession after a sudden rotation of the
 confining potential.
 The  equations describing the motion of the condensate
 are derived and the effects
 of superfluidity explicitly pointed out. The dependence 
 of the precession frequency on the relevant parameters of the problem is
 discussed. 
 The  proposed gyroscope is well suited to
 explore rotational effects 
 at the level of single quanta of circulation. 

\end{abstract}

 
 \narrowtext

After the realization of Bose-Einstein condensation \cite{jila}
 in magnetically trapped vapours of alkali atoms, cooled down to 
 extremely low temperatures, the superfluid behaviour of these systems 
has become  the object of   extensive experimental work.
This includes the study of  rotational 
properties, like quantized vortices \cite{jilav,ens} and the 
quenching of the moment of inertia \cite{oxford},
as well as the reduction of dissipative effects \cite{mitv}.

The purpose of this work is to show that Bose-Einstein condensed (BEC) gases
can be used to realize a quantum gyroscope 
where the effects of  superfluidity  show up in a very
peculiar way. 
Superfluid gyroscopes have been already realized with liquid $^4$He 
\cite{reppy} and $^3$He \cite{packard}
and have  been mainly used to investigate the 
nature of persistent currents in toroidal geometries. Most of experiments
with helium gyroscopes operate with many quanta of circulation.
Compared to liquid helium, 
trapped BEC gases are mesoscopic systems in the sense
 that the healing length, which provides a typical range of 
dynamic correlations and fixes the size of the vortex core,  is smaller, but 
not extremely
smaller than the size of the sample \cite{RMP}. For the same reason the angular 
momentum $N\hbar$ carried by  a single quantized vortex can have
visible effects on the global motion of the condensate as we will prove 
in this letter. 
A further feature that characterizes these systems is the very peculiar type of 
confinement which  yields  new possibilities for exploring
superfluid phenomena. Important gyroscopic  effects
associated with the occurrence of vortex lines have been already
 observed in trapped BEC gases \cite{ens2,jilap,jilap2}. 
 The authors of \cite{ens2}
have  succeeded in   testing
 the quantization of the 
angular momentum of  a single vortex line.
To this purpose one generates
 a quadrupole deformation 
in the plane orthogonal
to the vortex axis. The observed precession 
of the  deformation is  proportional
to the  angular momentum
carried by the vortex, in accordance with the predictions of 
theory \cite{ZS,sinha}. Notice that
in this experiment the  vortex line is 
not affected by the precession. 
The authors of \cite{jilap} and \cite{jilap2} have
instead observed the precession of a vortex line either displaced
or tilted from the symmetry axis of the condensate.
In this case the motion of the condensate is not 
affected at a macroscopic level and the precession involves the change
of the density on a more microscopic scale, of the order of
the  size of the vortex core. 

In the present work we  discuss the macroscopic precession of
the symmetry axis of a deformed condensate caused by the
sudden rotation of the 
trap, in the presence of a quantized vortex line. 
This precession corresponds to a full rotation of the condensate 
in 3D, which preserves the intrinsic shape of the system.
 
The geometry of the proposed  gyroscope is  illustrated in
fig.1. The system consists of  a dilute and cold  gas of  atoms with mass $m$
confined by an 
axi-symmetric trap of harmonic shape:
\begin{equation}
V_{ext} = {m\over 2}\left(\omega^2_{\perp}(x^2+y^2) + \omega^2_zz^2\right)
\label{Vext}
\end{equation}
where $\omega_{\perp}$ 
and $\omega_z$ are, respectively, 
the radial and axial trapping frequencies. The
trapping potential is characterized by the deformation
parameter 
\begin{equation}
\epsilon = {(\omega_{\perp}^2-\omega_z^2) 
\over (\omega_{\perp}^2+\omega_z^2)}
\label{epsilon}
\end{equation}
which  will be taken different from zero.
As a consequence of this anisotropy  
the atomic cloud is  also deformed. 
If the atoms are bosons, at very low temperatures they  undergo a phase transition 
associated with the macroscopic occupation of a single-particle state 
(Bose-Einstein condensation). This happens for temperatures below the critical
value $kT_c = 0.94 \hbar \omega_{ho}N^{1/3}$ where $\omega_{ho} 
= (\omega_z\omega^2_{\perp})^{1/3}$ is the geometrical average of the trapping
frequencies and $N$ is the number of atoms. Typical values of $T_c$ range
between $0.1$ and $1 \mu K$. Below $T_c$  
the system exhibits unique 
features which have been the object of systematic experimental 
and theoretical investigation.  For our purposes
the system should be very cold in order to minimize the dissipative 
effects produced by the thermal cloud which tend to destabilize
the vortex line \cite{Rok,gora}. For this reason we will 
 discuss here the
behaviour of the gas
at zero temperature. 

Let us  consider a Bose-Einstein condensate 
in equilibrium in the trap,
with no angular momentum, and let us suppose 
 that the symmetry axis of the trapping
potential is suddenly rotated 
in the $xz$ plane through an angle 
$\theta_0$.
After the rotation  the gas is
no longer in equilibrium and  starts oscillating around the new symmetry
axis. If the angle of rotation is small 
the gas will keep its intrinsic shape 
and the motion  will correspond to a periodic oscillation characterized
by the  angle 
$\theta(t)$, giving rise to the so-called scissors mode \cite{GS}. 
The motion 
is  very different from the one of a classical body because the inertia 
is strongly suppressed by superfluidity. In
particular the oscillation  frequency  
 does not vanish when the deformation (\ref{epsilon})
of the trap becomes small, 
as one would expect for a classical system.  
 
 The calculation of the   frequency of the scissors mode 
requires the solution of a non trivial  many-body problem where
the interactions among particles and the effects of quantum statistics 
play an important role.
An exact solution of the problem can be obtained
 for large samples, 
 such that the dimensionless parameter $Na/a_{ho}$ 
 is much larger than unity. This
limit corresponds to the so-called Thomas-Fermi (TF) regime and is 
well achieved in many available experimental configurations.
Here $a$ is the 
s-wave scattering length (of the order of $10^{-2}-10^{-3} \mu m$) while
$a_{ho}=\sqrt{\hbar/(m\omega_{ho})}$ is the oscillator length fixed by the 
geometrical average  of the 
trapping frequencies  
 (of the order of a few microns in standard magnetic traps). In the 
TF limit, the equilibrium density takes the form 
 of an inverted parabola: $n({\bf r}) = 
 m[\mu-V_{ext}({\bf r})]/(4\pi\hbar^2a)$ 
for $\mu \ge V_{ext}$ and zero elsewhere,   
where $\mu =(\omega_{ho}/2)(15Na/a_{ho})^{2/5}$ 
 is the chemical potential 
fixed by the normalization of the density. The
 surface where the density  vanishes defines an ellipsoid whose  
 principal radii
$R_{\perp}$ and $R_z$  are determined  by the relationship
  $\omega_{\perp}^2R^2_{\perp}= \omega_z^2R^2_{z} = 2\mu/m$.  
In the Thomas-Fermi limit 
the equations of motion take the simplified form of the 
hydrodynamic theory of superfluids \cite{pines}. With respect to the 
equations  of classical hydrodynamics they are  characterized but 
the additional  constraint of irrotationality.  It is worth noticing
that the "hydrodynamic" form of these equations is not the result
of collisional processes as happens in classical gases, but is the 
consequence of  superfluidity.
These equations admit an analytic solution
for the  scissors mode with frequency \cite{GS}
\begin{equation}
\omega_{scissors} = 
\sqrt{\omega_{\perp}^2 + \omega_z^2}
\label{omegascissors}
\end{equation}
This  mode has  recently been observed 
 \cite{oxford} confirming  the predictions 
of theory with high accuracy. 

Let us discuss now the case in which the condensate has initially
an intrinsic angular momentum. This can be due to a 
quantized vortex aligned along the symmetry axis of the condensate,
carrying one unit  $\hbar$ of angular momentum per particle. 
The size of the vortex core is fixed by the healing length
$\xi=(2m\mu/\hbar^2)^{-1/2}$ and is always smaller than the size of the
condensate.  
Also in the presence of the vortex 
the sudden rotation of the trap will excite the
scissors mode, but now the oscillation will precess
due to the  torque 
\begin{equation}
{d \over d t} <{\bf L}> = {1 \over i\hbar} <[{\bf L },H]> =-m <{\bf r \times 
\nabla}V_{ext}>
\label{torque}
\end{equation}
produced by the external  anisotropic potential (\ref{Vext}). Here
${\bf L}$ is the angular momentum operator and the expectation value is taken 
on the quantum mechanical state of the system. 
The value of the torque along the $y$ axis is $d<L_y>/dt = -m (\omega^2_{\perp}-
\omega_z^2)N<xz>$ and differs from zero immediately after the sudden rotation of 
the trap in the $xz$-plane, thereby causing the precession.

In order to derive the equation for the rotation of the 
condensate we notice that, if the  angle of rotation 
$\theta_0$ is  small compared to 
the deformation (\ref{epsilon}) of the trap, the shape of the sample is
preserved. Under this condition 
 the whole motion 
  can be described in terms of two angles: the 
inclination angle $\theta$ and the azimuthal angle $\phi$. This is a remarkable 
feature because  gases are highly compressible and their shape
can change very easily. 
 The values of $\theta$ and $\phi$ are
 related to the averages $<xz>$ and $<yz>$ by the 
geometrical relations
\begin{equation}
<xz> = {R^2\over 7} {\omega_z^2-\omega_{\perp}^2 \over 
\omega_z\omega_{\perp}}\theta\cos \phi \; ,\; \; \; \;  
<yz> = {R^2\over 7}{\omega_z^2-\omega_{\perp}^2 \over 
\omega_z\omega_{\perp}}\theta\sin \phi
\label{rotation}
\end{equation}
where  $R=\sqrt{R_{\perp} R_z}$  and we have assumed 
 $\theta \ll \epsilon$. 

 The crucial problem now is to obtain the equations for the averages $<xz>$ and $<yz>$ in the 
presence of the vortex.  This problem was  solved in \cite{ZS} 
 where it was shown 
that, in the Thomas-Fermi limit, 
the quadrupole operators $q_{\pm}=\sum_{k=1}^N(x_k\pm iy_k)z_k$
 excite, respectively,  just one collective  
mode $\mid \pm>$ with frequency $\omega_{\pm}$.
The solution of the Schr\"odinger equation,  in the presence
of a small perturbation generated by these operators, then takes the form
\begin{equation}
\mid \Psi(t)> =e^{-iE_0t/\hbar}
\left[ \mid 0> + c_+e^{-i\omega_+t}\mid +> + c_-e^{-i\omega_-t}\mid ->\right]
\label{psit}
\end{equation}
where $\mid0>$ is the equilibrium configuration, 
$\mid \pm>$ are eigenstates of the Hamiltonian corresponding 
to the elementary excitations
created by $q_{\pm}$ and the complex parameters $c_{\pm}$ 
characterize
the initial state of the system. One can also 
 show that the 
  strengths of the quadrupole operators $q_{\pm}$  are equal: 
$<0 \mid q_-q_+\mid 0> = <0 \mid q_+q_-\mid 0> $, while the 
splitting between the two frequencies  takes the simple form \cite{ZS}
\begin{equation}
\omega_+-\omega_- = {<l_z> \over m<x^2 + z^2>}
\label{splitting}
\end{equation}
In eq.(\ref{splitting}) the quantity
 $<\ell_z>$ is the angular momentum per particle 
in the equilibrium configuration, while
the average square radii  can be easily evaluated in
the Thomas-Fermi limit where one finds 
$<x^2> = <y^2>=  (2/7)\mu/m\omega^2_{\perp}$ and
$<z^2>  = (2/7)\mu/m\omega^2_{z} $. 
Result (\ref{splitting}) holds if  the splitting is 
small compared to  the 
unperturbed value (\ref{omegascissors}) of the collective frequency.  

The  scenario  emerging from the above discussion is 
quite clear. In the absence of vortices ($<l_z>=0$) the two modes
$\mid \pm>$  are degenerate 
and their frequency is  given by (\ref{omegascissors}). This is a simple
 consequence of  
time reversal symmetry. In the presence of the vortex there is  a lift of degeneracy which 
is determined by the angular momentum of the system according to 
(\ref{splitting}) \cite{note}.  
By expressing the average values $<xz>$ and $<yz>$ in terms of the expectation values
of the operators  $q_{\pm}$ and using explicitly 
eq. (\ref{psit}), one derives the following time
dependence for the angles $\theta$ and $\phi$:
\begin{equation}
\theta(t)\cos \phi(t) = \alpha\cos \omega_+t + \alpha^{\prime}\sin \omega_+t 
+\beta \cos \omega_-t + \beta^{\prime}\sin \omega_-t 
\label{solution1}
\end{equation}
\begin{equation}
\theta(t)\sin \phi(t) = -\alpha^{\prime}\cos \omega_+t + \alpha \sin \omega_+t
+\beta^{\prime} \cos \omega_-t - \beta\sin \omega_-t
\label{solution2}
\end{equation}
where the real coefficients $\alpha$, $\alpha^{\prime}$, 
$\beta$ and $\beta^{\prime}$
 depend on the initial conditions for $\theta$ and $\phi$. 
For example the normal modes $\mid \pm>$ can be separately excited by choosing 
$\alpha = \theta_0$, $\alpha^{\prime} = \beta = \beta^{\prime}=0$
and $\beta = \theta_0$, $\alpha = \alpha^{\prime}=\beta^{\prime}=0$ 
respectively. In both cases the inclination angle $\theta$ remains constant, 
$\theta(t)=\theta_0$, while the azimuthal angle  $\phi$ precesses according to 
the laws $\omega_+t$ and $-\omega_-t$, respectively.

In  the most relevant 
case of a sudden rotation in the $xz$ plane, the initial conditions
instead correspond to $\theta(0)=\theta_0, 
\theta^{\prime}(0)= \phi(0)=\phi^{\prime}(0) = 0$. In this case
the parameters of eqs.(\ref{solution1}-\ref{solution2}) 
are  $\alpha = \beta = \theta_0/2$,  $\alpha^{\prime}
=\beta^{\prime} = 0$
and the solutions take the form 
\begin{equation}
\phi(t) = {(\omega_+-\omega_-)t \over 2}
\label{phit}
\end{equation}
and 
\begin{equation}
\theta(t) = \theta_0\cos \left[{(\omega_++\omega_-)t \over 2}\right]
\label{thetat}
\end{equation} 
The simultaneous excitation of the $\mid \pm>$ modes 
is easily understood by noticing that  the rotation of the trap produces
a change in the external potential of the form $\delta V_{ext} =
m(\omega^2_{\perp}-\omega^2_z)\theta_0xz$. This gives rise to a perturbative term
in the Hamiltonian,  proportional to the combination
$(q_++q_-)$ of the quadrupole operators introduced above.
Equations (\ref{phit}) and (\ref{thetat})
show that the scissors oscillation, characterized by the
frequency $(\omega_++\omega_-)/2 \sim \omega_{scissors}$,  undergoes a precession $d\phi/dt$ 
 fixed by  the splitting (\ref{splitting}). 
 The relative precession  can be  expressed in the
form
\begin{equation}
{\omega_+-\omega_-\over \omega_++\omega_-} = {7\over 2} {<\ell_z>\over \hbar}
{\lambda^{5/3}
\over (1+\lambda^2)^{3/2}}\left(15N{a\over a_{ho}}\right)^{-2/5}
\label{ratio}
\end{equation}
where 
$\lambda = \omega_z/\omega_{\perp} = \sqrt{(1-\epsilon)/(1+\epsilon)}$ and
we have 
used the Thomas-Fermi result for $<x^2>$ and $<z^2>$. 
The precession frequency $d\phi/dt$, and hence the ratio (\ref{ratio}),
depends explictly on the value of the angular momentum per particle
which, in the case of a 
single quantized vortex aligned along the symmetry axis of the condensate,
is given by $<\ell_z> = \hbar$. 
The ratio (\ref{ratio})  depends  also 
on the shape of   the trap. 
This provides further
flexibility to optimize   the  visibility of the precession.
In typical experimental configurations   the Thomas-Fermi
combination  $(15Na/a_{ho})^{-2/5}$ is of the 
order of $10^{-2}$ so that for a 
highly elongated
trap ($\lambda \ll 1$) the relative precession is small. For 
values of $\lambda$ closer to unity 
the ratio (\ref{ratio}) becomes larger. 
With suitable choices of the parameters of the trap the precession
can be easily of the order of 1 Hz and should be 
consequently observable by imaging the atomic cloud at different
times. 
  
In conclusion we have shown that Bose-Einstein condensed gases confined
in harmonic traps  can be used
to realize a quantum gyroscope characterized 
by two important superfluid effects: the reduced
value of the inertia of the sample  
 and the quantization of the 
 angular momentum associated with the vortex. 
 The proposed gyroscope is characterized by the 
precession of the symmetry axis of the condensate around the symmetry axis of the confining
trap, the precession frequency being fixed by the angular momentum carried
by the vortex line. The  
experimental realization of the proposed gyroscope
would provide a further  tool to explore the  intriguing
features exhibited by rotating Bose-Einstein condensed gases, including the
stability \cite{fetter,perez} and the life time \cite{gora} of  vortex lines.

It is a pleasure to thank E. Cornell, 
J. Dalibard, A. Fetter, R. Packard, J. Reppy, E. Varauqaux 
and J. Vinen for useful comments and discussions. I also like to thank the hospitality 
of the Lorentz Center in Leiden and of the Newton Institute in Cambridge. 
This research is supported 
by the Ministero della Ricerca Scientifica e Tecnologica (MURST).

\noindent
FIGURE CAPTION:

\noindent
{Schematic picture of the BEC gyroscope. The condensate is suddenly rotated 
with respect to the symmetry axis of the trap ($z$-th axis) and exhibits
precession in the presence of a quantized vortex line.}


\begin{references}
\bibitem{note} A similar scenario takes place if we excite 
the system with the quadrupole
operators $\sum_{k=1}^N(x_k\pm i y_k)^2$.
 This choice corresponds to generating a quadrupole
deformation in the plane perpendicular the symmetry axis, like 
in the experiment of \cite{ens2}. In this case the unperturbed
frequency of the quadrupole oscillation 
is $\sqrt{2}\omega_{\perp}$, while the 
splitting takes the form $\omega_+-\omega_- = 
(2/m)<\ell_z>/(<x^2>+<y^2>)$ \cite{ZS}.
\end{references}
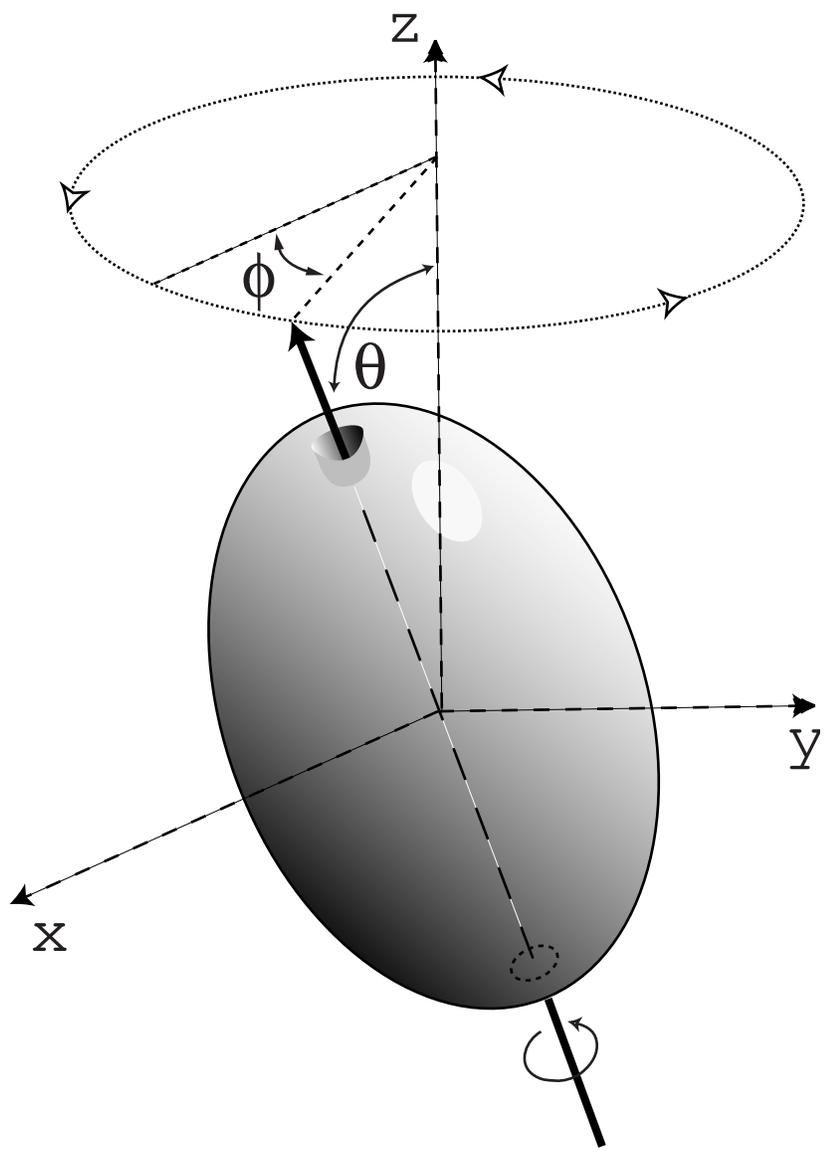
\end{document}